\documentclass[11pt, aps, nofootinbib]{revtex4-1}

\usepackage{geometry}
\usepackage{latexsym}
\usepackage{graphicx}
\usepackage{amsmath}
\usepackage{amssymb}
\usepackage{color}
\usepackage{mathtools}
\usepackage[colorlinks=true,a4paper=true,backref=false,linktocpage=true,citecolor=blue,urlcolor=blue,linkcolor=blue,pdfpagemode=UseOutlines]{hyperref}

\newcommand{\be}{\begin{equation}}
\newcommand{\ee}{\end{equation}}
\newcommand{\bea}{\begin{eqnarray}}
\newcommand{\eea}{\end{eqnarray}}

\newcommand{\LaSapienza}{Physics Department and INFN Sezione di Roma La Sapienza,\\ Piazzale Aldo Moro 5, 00185 Roma, Italy}
\newcommand{\RomatreINFN}{Istituto Nazionale di Fisica Nucleare, Sezione di Roma Tre,\\ Via della Vasca Navale 84, I-00146 Rome, Italy}
\newcommand{\SNS}{Scuola Normale Superiore, Piazza dei Cavalieri 7, I-56126, Pisa, Italy}
\newcommand{\INFNpisa}{Istituto Nazionale di Fisica Nucleare, Sezione di Pisa,\\ Largo Bruno Pontecorvo 3, I-56127 Pisa, Italy}

\begin{document}

\title{Exclusive determinations of $\vert V_{cb} \vert$ and $R(D^{*})$ through unitarity}

\author{G.~Martinelli}\affiliation{\LaSapienza}
\author{S.~Simula}\affiliation{\RomatreINFN}
\author{L.~Vittorio}\affiliation{\SNS},\affiliation{\INFNpisa}

\begin{abstract}
In this work we apply the Dispersive Matrix (DM) method of Refs.~\cite{DiCarlo:2021dzg,Martinelli:2021onb} to the lattice computations of the Form Factors (FFs) entering the semileptonic $B \to D^* \ell \nu_\ell$ decays, recently produced by the FNAL/MILC Collaborations~\cite{FermilabLattice:2021cdg} at small, but non-vanishing values of the recoil variable ($w-1$). Thanks to the DM method we obtain the FFs in the whole kinematical range accessible to the decay in a completely model-independent and non-perturbative way, implementing exactly both unitarity and kinematical constraints. Using our theoretical bands of the FFs we extract $\vert V_{cb} \vert$ from the experimental data and compute the theoretical value of $R(D^*)$. Our final result for $\vert V_{cb} \vert$ reads $\vert V_{cb} \vert = (41.3 \pm 1.7) \cdot 10^{-3}$, compatible with the most recent inclusive estimate at the $0.5\sigma$ level. Moreover, we obtain the pure theoretical value $R(D^*) = 0.275 \pm 0.008$, which is compatible with the experimental world average at the $\sim 1.3 \sigma$ level.
\end{abstract}

\maketitle

\newpage

\section{Introduction}
\label{sec:introduction}

The exclusive semileptonic $B \to D^* \ell \nu_\ell$ decay is a very intriguing process from a phenomenological point of view, mainly for two reasons: the first one is the $\vert V_{cb} \vert$ puzzle, $i.e.$, the tension between the inclusive and exclusive determinations of the Cabibbo-Kobayashi-Maskawa (CKM) matrix element $|V_{cb}|$; the second one is the discrepancy between the theory and the experiments in the determination of the $\tau / \mu$ ratio of the branching fractions, $R(D^*)$, which is a fundamental test of Lepton Flavour Universality in the Standard Model.

In this work we determine $\vert V_{cb} \vert$ and $R(D^*)$ using the final lattice results of the FFs entering the semileptonic $B \to D^*$ decays, produced recently by the FNAL/MILC Collaborations~\cite{FermilabLattice:2021cdg}. 
To this end we adopt the DM method of Refs.~\cite{DiCarlo:2021dzg,Martinelli:2021onb}, originally proposed in Ref.\,\cite{Lellouch:1995yv}, to obtain the FFs in the whole kinematical range starting from the lattice computations performed at small, but non-vanishing values of the recoil variable ($w-1$). 
The crucial advantage of the DM approach is that the extrapolations of the FFs can be performed in a fully model-independent and non-perturbative way, since no assumption about the functional dependence of the FFs on the recoil $w$ is made and all the theoretical inputs of the DM approach are computed on the lattice (see Refs.~\cite{Martinelli:2021frl,Martinelli:2021onb}). 
Then, as in Ref.~\cite{Martinelli:2021onb}, we analyse the experimental data by performing a bin-per-bin extraction of $\vert V_{cb} \vert$ using the DM bands of the FFs.

Our result is $\vert V_{cb} \vert = (41.3 \pm 1.7) \cdot 10^{-3}$, which is compatible with the most recent inclusive determination $\vert V_{cb} \vert_{\rm{incl}} = (42.16 \pm 0.50) \cdot 10^{-3}$~\cite{Bordone:2021oof}. 
This implies that the exclusive and the inclusive determinations of $\vert V_{cb} \vert$ are now compatible at the $0.5\sigma$ level.
Note that using other weak processes a similar indication was already claimed by the UT\emph{fit} Collaboration in Ref.~\cite{Alpigiani:2017lpj} and more recently in Ref.~\cite{Buras:2021nns}.

The uncertainty obtained for $\vert V_{cb} \vert$ is larger than those of other analyses available in the literature (see, e.g., Refs.~\cite{Bigi:2016mdz,Jaiswal:2017rve, FlavourLatticeAveragingGroup:2019iem}). In this respect we stress that in our procedure the lattice and the experimental data are two sources of information that we always kept separate at variance with the other analyses. To be more precise, the lattice computations are used in order to derive the allowed bands of the FFs, while the experimental measurements are considered only for the final determination of $\vert V_{cb} \vert$, avoiding in this way any possible bias of the experimental distribution on the theoretical predictions and hence on the extracted value of $\vert V_{cb}\vert$. This difference justifies the larger uncertainty of our estimate of $\vert V_{cb} \vert$ with respect to those of the other analyses.

The DM method allows also to predict the ratio $R(D^{*})$ from theory, obtaining $R(D^*) = 0.275 \pm 0.008$, which is compatible with the experimental world average $R(D^*) = 0.295 \pm 0.011 \pm 0.008$~\cite{HFLAV:2019otj} at the $\sim 1.3 \sigma$ level.

\section{The unitarity bands of the Form Factors}
\label{sec:FFs}

We apply the DM method to the final lattice computations of the FFs provided by the FNAL/MILC Collaborations~\cite{FermilabLattice:2021cdg}. There, in the ancillary files, the authors give the synthetic values of the FFs $g(w), f(w), \mathcal{F}_1(w)$ and $\mathcal{F}_2(w)$ at three non-zero values of the recoil variable ($w-1$), namely $w = \{1.03,1.10,1.17\}$, together with their correlations. 
In what follows, we will refer to the pseudoscalar FF $P_1(w)$, which is connected to $\mathcal{F}_2(w)$ through the relation $P_1(w) =  \mathcal{F}_2(w) \sqrt{r} / (1 + r)$, where $r \equiv m_{D^*}/m_B \simeq 0.38$. 

A brief description of the main features of the DM approach is given in Appendix\,\ref{sec:appA}, while the nonperturbative values of the relevant susceptibilities, computed on the lattice in Ref.~\cite{Martinelli:2021frl}, are collected in Appendix\,\ref{sec:appB}.

\begin{figure}[htb!]
\begin{center}
\includegraphics[scale=0.625]{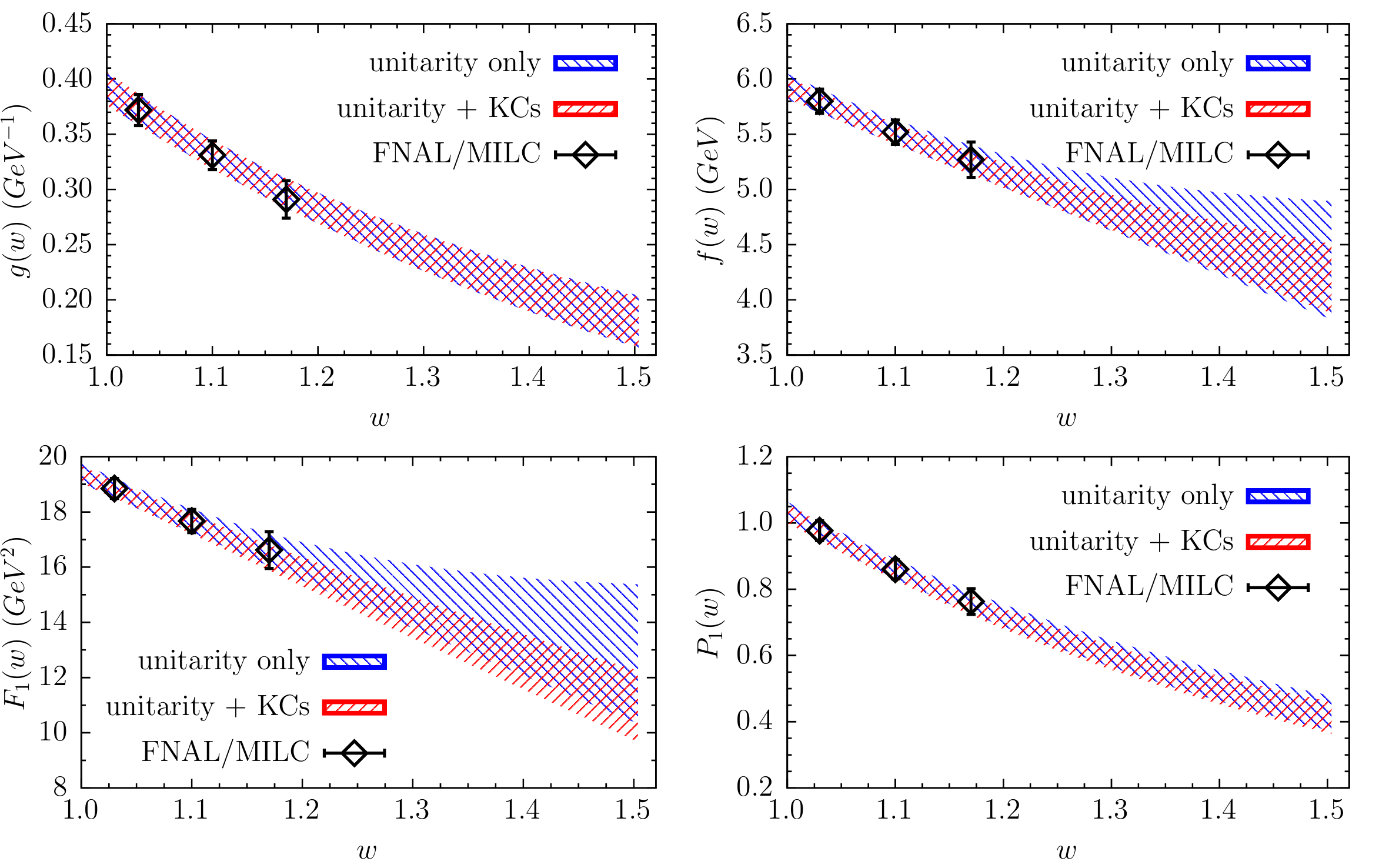}
\caption{\it \small The bands of the FFs $g(w)$, $f(w)$, $\mathcal{F}_1(w)$ and $P_1(w)$ computed by the DM method after imposing either the unitarity filters only (bue bands) or the unitarity filters and the two KCs~(\ref{eq:KC1})-(\ref{eq:KC2}) (red bands). The FNAL/MILC values~\cite{FermilabLattice:2021cdg} used as inputs for the DM method are represented by the black diamonds.}
\label{fig:FFs}
\end{center}
\end{figure}

Using multivariate Gaussian distributions we generate a sample of $10^5$ events, each of which is composed by 12 data points for the FFs (3 points for each FF) and 3 data points for the relevant susceptibilities.
Then, we apply the three unitarity filters of the DM method to the FFs $g$, $(f, \mathcal{F}_1)$ and $P_1$ (see Eqs.\,(\ref{eq:UTfilter1})-(\ref{eq:UTfilter3}) of Appendix\,\ref{sec:appB}).
They are satisfied only by a reduced number of events. Indeed, the percentage of the surviving events turns out to be only $\approx 1\%$ after imposing the three unitarity constraints on  $g$, $(f, \mathcal{F}_1)$ and $P_1$. 
The subset of the surviving events turns out to be very well approximated by a Gaussian Ansatz. Thus, on such subset we recalculate the mean values, uncertainties and correlations of the FFs (and susceptibilties).
The changes in the mean values and uncertainties turn out to be quite small, while the application of the unitarity filters has its major impact on the correlations among the FFs.
We repeat the generation of the sample using the new input values and we apply again the unitarity filters. 
Adopting the above {\it iterative procedure} the fraction of surviving events for all the FFs increases each time reaching $\simeq 79 \%$ already after two iterations and $\simeq 91 \%$ after three iterations.
We stop the iteration procedure when the changes of the mean values, uncertainties and correlations of the FFs (and susceptibilties) are less than few permil, which occurs in practice after five iterations\footnote{In Ref.\,\cite{Martinelli:2021onb} we adopted a skeptical approach, inspired to the results of Refs.\,\cite{DAgostini:2020vsk,DAgostini:2020pim}, as a tool for generating a larger set of events, among which one can search for those passing the given constraint, obtaining in this way samples with a statistically significant number of events. The practical implementation of the skeptical approach is quite time-consuming, while with the iterative procedure one can easily obtain samples with almost all events passing the given filter and normally distributed. The iterative procedure is simpler and very effective.}.
The resulting DM bands of the FFs are shown in the whole range of values of the recoil $w$ as the blue bands in Fig.~\ref{fig:FFs}.

We have also to impose two kinematical constraints (KCs) that relate the FFs $f$ and $\mathcal{F}_1$ at $w = 1$ and the FFs $\mathcal{F}_1$ and $P_1$ at $w = w_{max} = (1 + r^2) / (2r) \simeq 1.50$, namely\footnote{The KC at $w = 1$ comes from the fact that at zero recoil only two out of the three helicity amplitudes for the final $D^*$-meson at rest are independent. The second KC is related to the cancellation of any apparent kinematical singularity at $q^2 = 0$ (i.e., $w = w_{max}$) in the Lorentz decomposition of the matrix elements of the $(V - A)$ weak current.} 
\bea
    \label{eq:KC1}
    \mathcal{F}_1(1) & = & m_B (1 - r) f(1) ~ , ~ \\[2mm]
    \label{eq:KC2}
     P_1(w_{max}) & = & \frac{\mathcal{F}_1(w_{max})}{m_B^2 (1 + w_{max}) (1 -r) \sqrt{r}} ~ . ~
\eea
The procedure adopted to include, e.g., the KC\,(\ref{eq:KC1}) is illustrated in details in the Appendix\,\ref{sec:appC}.
We apply again the iterative procedure to increase each time the percentage of surviving events after imposing the filters corresponding to the two KCs~(\ref{eq:KC1})-(\ref{eq:KC2}). 
We require a fraction of surviving events $\gtrsim 96 \%$ after imposing all the filters (the three unitarity and the two KC filters).
The resulting DM bands of the FFs are shown as the red bands in Fig.~\ref{fig:FFs}. 
It can be seen that the proper inclusion of the KCs (particularly the one at $w = w_{max}$) has a crucial impact on the extrapolation of the FFs $f$ and $\mathcal{F}_1$ at large values of the recoil $w$.
The extrapolations of the FFs at $w  = w_{max}$ read 
\bea
    \label{eq:f_wmax}
    f(w_{max}) & = & 4.19 \pm 0.31 ~ \mbox{GeV} ~ , ~ \nonumber \\[2mm]
    \label{eq:g_wmax}
    g(w_{max}) &= & 0.180 \pm 0.023  ~ \mbox{GeV}^{-1} ~ , ~ \\[2mm]
    \label{eq:F1_wmax}
    \mathcal{F}_1(w_{max}) & = & 11.0 \pm 1.3  ~ \mbox{GeV}^2 ~ , ~ \nonumber \\[2mm]
    \label{eq:P1_wmax}
    P_1(w_{max}) & = & 0.411 \pm 0.048 ~ . ~ \nonumber
\eea

\section{Determination of $\mathbf{\vert V_{cb} \vert}$}
\label{sec:Vcb}

We start from the measurements of the differential decay widths performed by the Belle Collaboration for the semileptonic $B \to D^* \ell \nu_\ell$ decays~\cite{Abdesselam:2017kjf, Waheed:2018djm}\footnote{We do not make use of the results of the BaBar Collaboration given recently in Ref.~\cite{BaBar:2019vpl}, since only synthetic data based on parameterization-dependent fits of the FFs are available.}.
We now determine a new exclusive estimate of $\vert V_{cb} \vert$ by performing a bin-per-bin study of the Belle experimental data. 
The latter ones are given in the form of 10-bins distribution of the quantity $d\Gamma / dx$, where $x$ is one of the four kinematical variables of interest ($x = w, \cos \theta_l, \cos \theta_v, \chi$) (see \cite{Martinelli:2021onb} for the expressions of the four-dimensional differential decay widths and Refs.~\cite{Abdesselam:2017kjf, Waheed:2018djm} for the specific values of the four variables $x$ in each bin). 
First of all, we generate a sample of values of the FFs $g$, $f$, $\mathcal{F}_1$ and $P_1$ to be used for each of the experimental bins using the DM method described in the previous Section. With such FFs we compute the theoretical predictions $d\Gamma^{th} / dx / |V_{cb}|^2$ for each experimental bin. 
We also generate an independent sample of values of the experimental differential decay widths $d\Gamma^{exp} / dx$ for all the bins. 
For each event of the sample we compute $|V_{cb}|$ as the square root of the ratio of the experimental over the theoretical differential decay widths for all the bins.
Using the produced events for $|V_{cb}|$, whose distribution turns out to be very well approximated by a Gaussian Ansatz, we compute the mean values $\vert V_{cb} \vert_i$  and the corresponding covariance matrix $\mathbf{C}_{ij}$ for all the bins ($i, j = 1, ..., 10$). 
Finally, adopting the best constant fit over the 10 bins we compute $|V_{cb}|$ and its variance $\sigma^2_{\vert V_{cb} \vert} $ for each of the four kinematical variables and for each of the two experiments~\cite{Abdesselam:2017kjf, Waheed:2018djm} as 
\bea
    \label{eq:muVcb}
    \vert V_{cb} \vert & = & \frac{\sum_{i, j=1}^{10} (\mathbf{C}^{-1})_{ij} \vert V_{cb} \vert_j}{\sum_{i, j=1}^{10} (\mathbf{C}^{-1})_{ij}} ~ , ~ \\[2mm]
    \label{eq:sigmaVcb}
    \sigma^2_{\vert V_{cb} \vert} & = & \frac{1}{\sum_{i, j=1}^{10} (\mathbf{C}^{-1})_{ij}} ~ . ~
\eea

\begin{figure}[htb!]
\begin{center}
\includegraphics[scale=0.625]{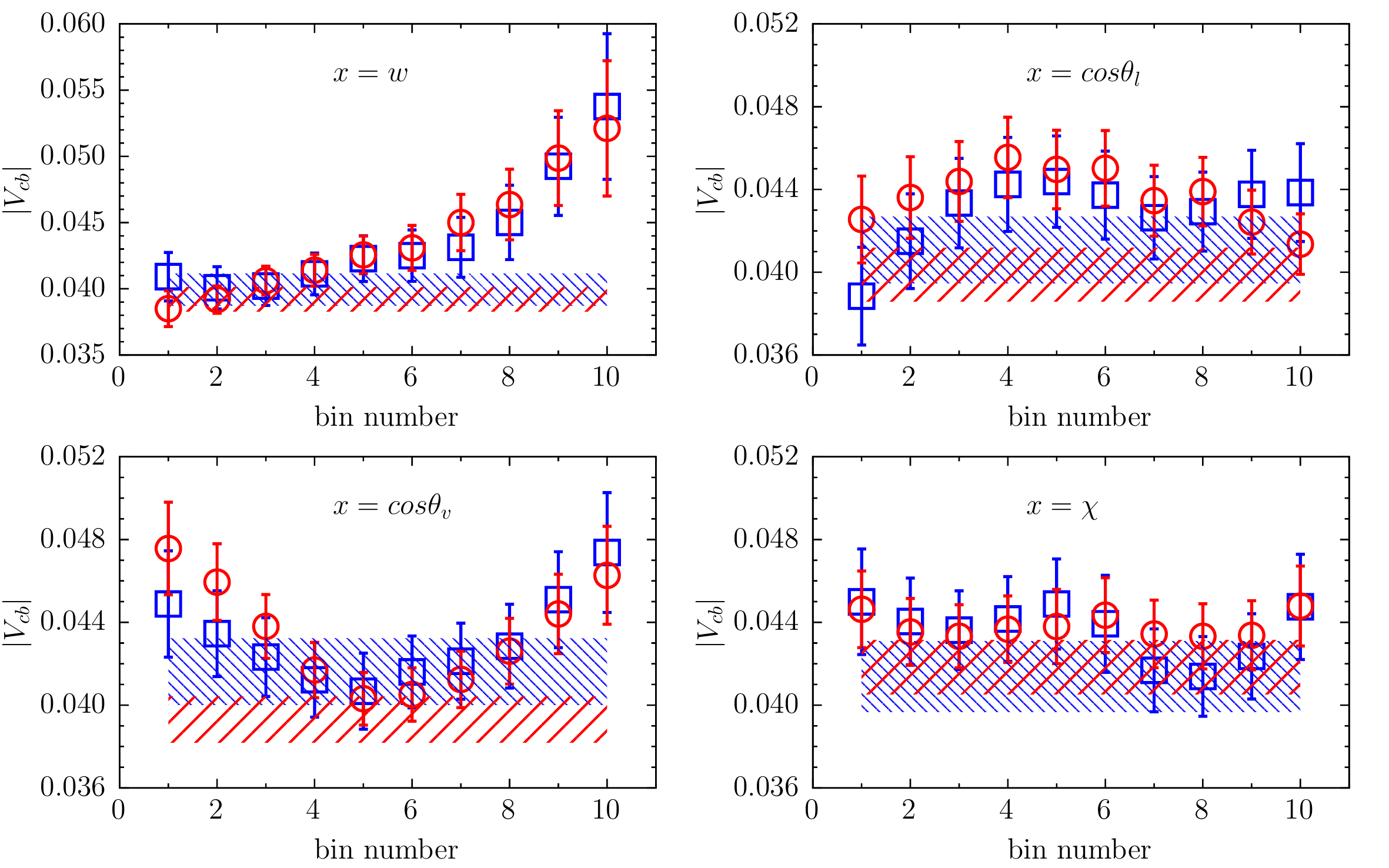}
\caption{\it \small The bin-per-bin estimates of $\vert V_{cb} \vert$ and their weighted means (\ref{eq:muVcb})-(\ref{eq:sigmaVcb}) for each kinematical variable $x$ and for each experiment adopting the original covariance matrices of the Belle experiments. The blue squares and the red circles correspond respectively to the first~\cite{Abdesselam:2017kjf} and to the second~\cite{Waheed:2018djm} set of the Belle measurements. The dashed blue (red) bands are the results of Eqs.~(\ref{eq:muVcb})-(\ref{eq:sigmaVcb}) in the case of the blue squares (red circles) for each variable $x$ (see Refs.~\cite{Abdesselam:2017kjf, Waheed:2018djm} for the specific values of the four variables $x$ in each bin).}
\label{fig:Vcb_orig}
\end{center}
\end{figure}

In Fig.~\ref{fig:Vcb_orig} we show the bin-per-bin distributions of $\vert V_{cb} \vert$ for each kinematical variable $x$ and for each experiment, together with their final weighted mean values. The latter ones are collected also in Table~\ref{tab:Vcb_orig} together with the corresponding values of the reduced $\chi^2$-variable, $\chi^2/(\mbox{d.o.f.})$, being the number of d.o.f.~equal to 9.
As already noted in Ref.~\cite{Martinelli:2021onb}, we observe anomalous underestimates of the mean values of $\vert V_{cb} \vert$ in the case of some of the variables $x$, which correspond also to large values of the reduced $\chi^2$-variable.

\begin{table}[htb!]
\renewcommand{\arraystretch}{1.5}
\begin{center}
\begin{tabular}{||c||c|c|c|c||}
\hline \hline
 experiment & ~$\vert V_{cb} \vert(x = w)$~ & ~$\vert V_{cb} \vert(x = \mbox{cos}\theta_l)$~ & ~$\vert V_{cb} \vert(x = \mbox{cos}\theta_v)$~ & ~$\vert V_{cb} \vert(x = \chi)$~\\
\hline \hline
Ref.~\cite{Abdesselam:2017kjf} & ~0.0399~(12)~ & ~0.0411~(16)~ & ~0.0416~(16)~ & ~0.0414~(17)~\\
  $\chi^2/(\mbox{d.o.f.})$            & ~1.72~ &  ~1.10~ &~1.21~ &~1.45~ \\
\hline \hline
Ref.~\cite{Waheed:2018djm}     & ~0.0392~~(9)~ & ~0.0399~(13)~ & ~0.0393~(11)~ & ~0.0418~(13)~\\
$\chi^2/(\mbox{d.o.f.})$              & ~1.62~ &  ~2.41~ &~3.77~ &~0.79~ \\
\hline \hline
\end{tabular}
\end{center}
\renewcommand{\arraystretch}{1.0}
\caption{\it \small Mean values and uncertainties of the CKM element $\vert V_{cb} \vert$ obtained by the correlated average~(\ref{eq:muVcb})-(\ref{eq:sigmaVcb}) for each of the four kinematical variables $x$ and for each of the two experiments~\cite{Abdesselam:2017kjf,Waheed:2018djm} adopting the original covariance matrices  of the Belle experiments. The corresponding values of the reduced $\chi^2$-variable, $\chi^2/(\mbox{d.o.f.})$, are also shown.}
\label{tab:Vcb_orig}
\end{table}

We adopt the alternative strategy described in Ref.~\cite{Martinelli:2021onb} for each of the two Belle experiments. 
We consider the relative differential decay rates given by the ratios $(d\Gamma / dx) / \Gamma$ (where $x = w,  \cos \theta_l, \cos \theta_v, \chi$) for each bin by using the experimental data. 
In this way we guarantee that the sum over the bins is exactly independent (event by event) of the choice of the variable $x$. 
Hence, we compute a \emph{new} correlation matrix using the events for the ratios $(d\Gamma / dx) / \Gamma$. 
The new correlation matrix has four eigenvalues equal to zero, because the sum over the bins of each of the four variable $x$ is always equal to unity. In other words, the number of independent bins for the ratios is 36 and not 40 for each experiment. 
Then, following Ref.~\cite{Martinelli:2021onb} a \emph{new} covariance matrix of the experimental data is constructed by multiplying the new correlation matrix by the original uncertainties associated to the measurements.

\begin{figure}[htb!]
\begin{center}
\includegraphics[scale=0.625]{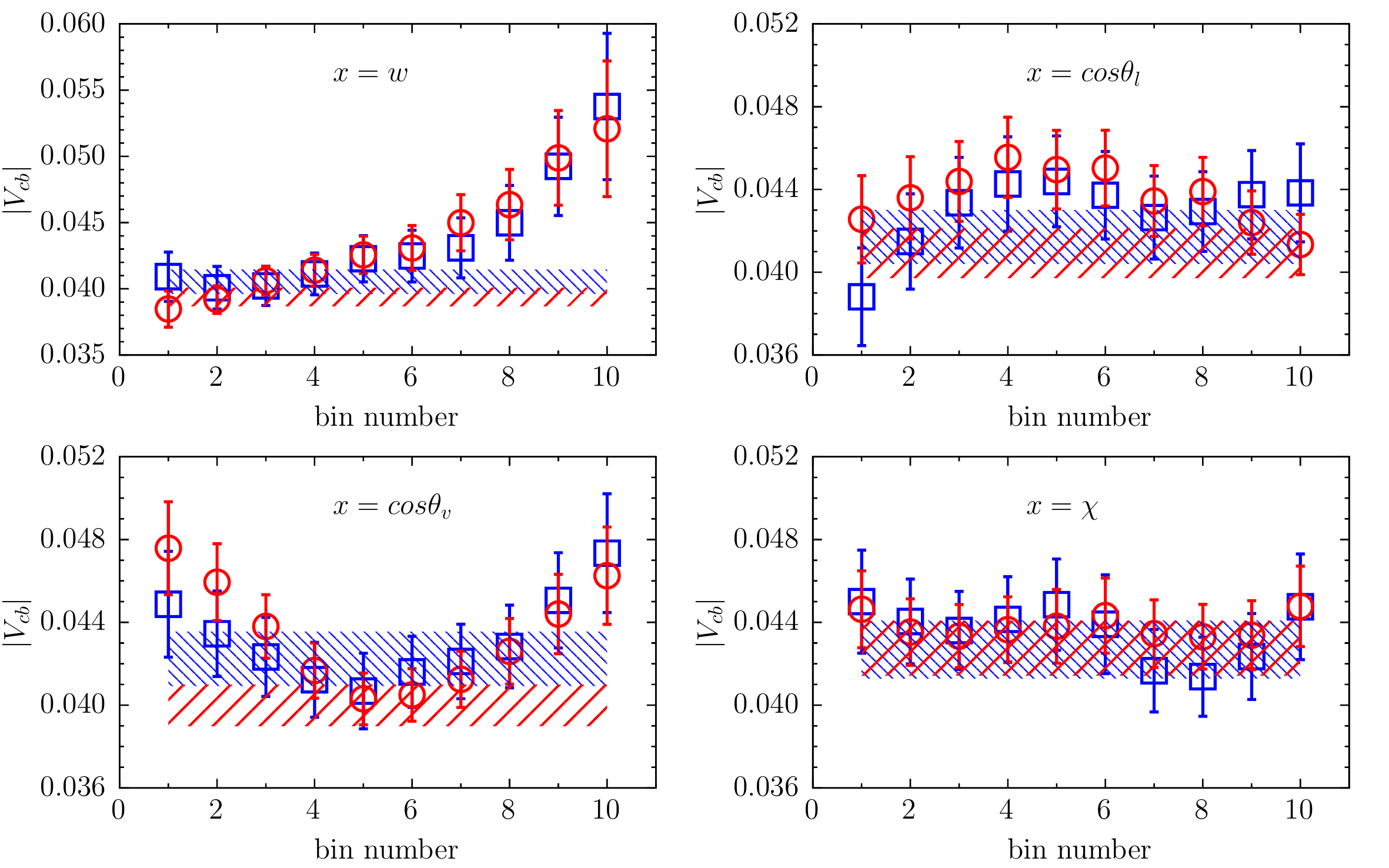}
\caption{\it \small The same as in Fig.~\ref{fig:Vcb_orig}, but using the new experimental covariance matrices described in the text.}
\label{fig:Vcb_new}
\end{center}
\end{figure}

Thus, we repeat the whole procedure for the extraction of $\vert V_{cb} \vert$ using the new experimental covariance matrices. 
In Fig.~\ref{fig:Vcb_new} we show the bin-per-bin distributions of $\vert V_{cb} \vert$ for each kinematical variable $x$ and for each experiment, together with their final weighted mean values. 
The latter ones are collected also in Table~\ref{tab:Vcb_new}.
A drastic improvement of the values of the reduced $\chi^2$-variable is obtained for each of the kinematical variable $x$ and for each of the two Belle experiments.

\begin{table}[htb!]
\renewcommand{\arraystretch}{1.5}
\begin{center}
\begin{tabular}{||c||c|c|c|c||}
\hline \hline
 experiment & ~$\vert V_{cb} \vert(x = w)$~ & ~$\vert V_{cb} \vert(x = \mbox{cos}\theta_l)$~ & ~$\vert V_{cb} \vert(x = \mbox{cos}\theta_v)$~ & ~$\vert V_{cb} \vert(x = \chi)$~\\
\hline \hline
Ref.~\cite{Abdesselam:2017kjf} & ~0.0405~(9)~ & ~0.0417~(13)~ & ~0.0422~(13)~ & ~0.0427~(14)~\\
  $\chi^2/(\mbox{d.o.f.})$            & ~1.01~ &  ~0.89~ &~0.66~ &~0.72~ \\
\hline \hline
Ref.~\cite{Waheed:2018djm}     & ~0.0394~(7)~ & ~0.0409~(12)~ & ~0.0400~(10)~ & ~0.0427~(13)~\\
$\chi^2/(\mbox{d.o.f.})$              & ~1.21~ &  ~1.36~ &~1.99~ &~0.38~ \\
\hline \hline
\end{tabular}
\end{center}
\renewcommand{\arraystretch}{1.0}
\caption{\it \small The same as in Table~\ref{tab:Vcb_orig}, but using the new experimental covariance matrices described in the text.}
\label{tab:Vcb_new}
\end{table}

Then, we combine the mean values of Table~\ref{tab:Vcb_new} through the formul\ae\,\cite{EuropeanTwistedMass:2014osg}
\bea
    \label{eq:average28}
    \mu_x & = & \frac{1}{N} \sum_{k=1}^N x_k ~ , ~ \\[2mm]
    \label{eq:sigma28}
    \sigma_x^2 & = & \frac{1}{N} \sum_{k=1}^N \sigma_k^2 + \frac{1}{N} \sum_{k=1}^N(x_k-\mu_{x})^2 ~ , ~
\eea
where the second term in the r.h.s.~of Eq.~(\ref{eq:sigma28}) accounts for the spread of the values of $\vert V_{cb} \vert$ corresponding to the various kinematical variables and experiments.
We obtain for each of the two Belle experiments the averages 
\bea
    \label{eq:Vcb}
    \vert V_{cb} \vert & = & (41.8 \pm 1.5) \cdot 10^{-3} \qquad \mbox{Ref.~\cite{Abdesselam:2017kjf}} ~ \nonumber \\[2mm]
                               & = & (40.8 \pm 1.7) \cdot 10^{-3} \qquad \mbox{Ref.~\cite{Waheed:2018djm}}~ \nonumber    
\eea
and by further combining the two Belle experiments the final estimate
\be
    \label{eq:Vcb_final}
    \vert V_{cb} \vert = (41.3 \pm 1.7) \cdot 10^{-3} ~ , ~
\ee
which is compatible with the most recent inclusive determination $\vert V_{cb} \vert_{\rm{incl}} = (42.16 \pm 0.50) \cdot 10^{-3}$~\cite{Bordone:2021oof} at the $0.5\sigma$ level. 

Without the modification of the experimental covariance matrices (i.e.\,using the eight mean values shown in Table~\ref{tab:Vcb_orig} and in Fig.~\ref{fig:Vcb_orig}) the final estimate of $\vert V_{cb} \vert$ would have read
\be
    \vert V_{cb} \vert = (40.5 \pm 1.7) \cdot 10^{-3} ~ , ~ \nonumber
\ee
which is still compatible with the most recent inclusive determination at the $1\sigma$ level.

From Fig.~\ref{fig:Vcb_new} we note that:
\begin{itemize}

\item[a)] in the top left panel the value of $ \vert V_{cb} \vert$ exhibits some dependence on the specific $w$-bin. The value obtained adopting a constant fit is dominated by the bins at small values of the recoil, where direct lattice data are available and the lenght of the momentum extrapolation is limited;

\vspace{0.5cm}

\item[b)] in the bottom left panel the value of $ \vert V_{cb} \vert$ deviates from a constant behavior, as it is also signaled by the large value of the corresponding reduced $\chi^2$-variable for the second set of the Belle measurements. Instead of a constant fit, we try a quadratic one of the form $\vert V_{cb} \vert \left[1 + \delta B \, \mbox{cos}^2(\theta_v) \right]$, suggested by the structure of the differential decay rate $d\Gamma / d \mbox{cos}(\theta_v)$ within the Standard Model and beyond (see, e.g., Ref.\,\cite{Ivanov:2016qtw})). If the dependence of the experimental and theoretical decay rates upon $\mbox{cos}(\theta_v)$ were the same, then the parameter $ \delta B$ would identically vanish. Instead we get a non-vanishing value of the parameter $ \delta B$, namely: $\vert V_{cb} \vert = (41.1 \pm 1.4) \cdot 10^{-3}$ and $ \delta B = 0.144 \pm 0.074$ for the first set~\cite{Abdesselam:2017kjf} of the Belle measurements, and $\vert V_{cb} \vert = (40.8 \pm 1.0) \cdot 10^{-3}$ and $ \delta B = 0.184 \pm 0.050$ for the second set~\cite{Waheed:2018djm}. The values of  $\vert V_{cb} \vert $ are consistent with each other and also with the corresponding values obtained adopting a constant fit and shown in the fourth column of Table~\ref{tab:Vcb_new}.
\end{itemize}

\begin{figure}[htb!]
\begin{center}
\includegraphics[scale=0.50]{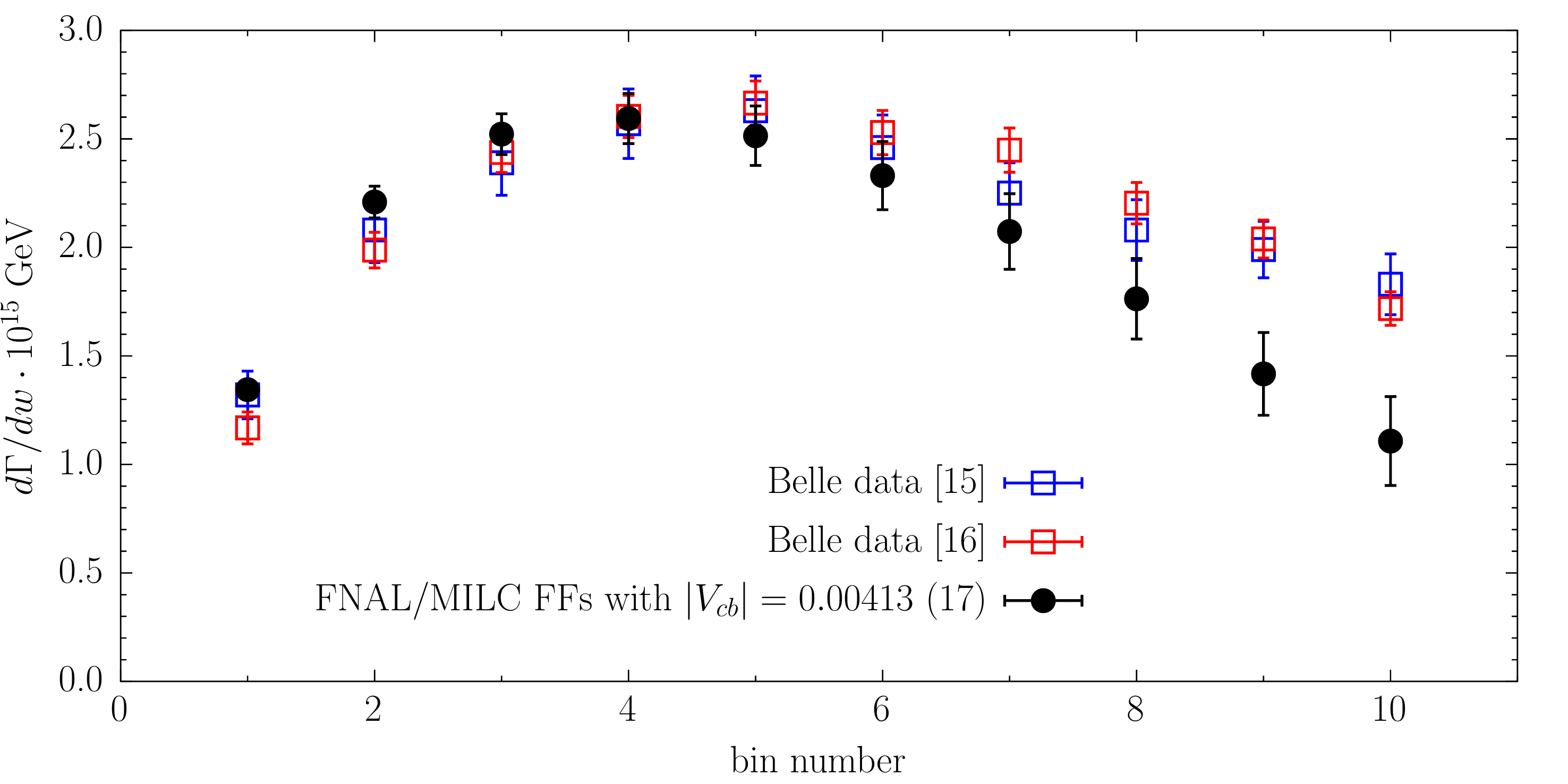}
\caption{\it \small The differential decay width $d\Gamma / dw$ measured by the two Belle experiments\,\cite{Abdesselam:2017kjf, Waheed:2018djm} in the 10 experimental $w$-bins compared with the corresponding theoretical predictions obtained in this work using the DM bands for the FFs, based on the FNAL/MILC synthetic data\,\cite{FermilabLattice:2021cdg} (see Fig.\,\ref{fig:FFs}), and adopting for $|V_{cb}|$ our final determination given by Eq.\,(\ref{eq:Vcb_final}).}
\label{fig:comparison}
\end{center}
\end{figure}

Both observations may be related to a different $w$-slope of the theoretical FFs based on the lattice results of Ref.~\cite{FermilabLattice:2021cdg} with respect to the Belle experimental data~\cite{Abdesselam:2017kjf,Waheed:2018djm}, as shown in Fig.\,\ref{fig:comparison}. This crucial issue (a kind of a new {\it slope puzzle}) needs to be further investigated by forthcoming calculations of the FFs at non-zero recoil expected from the JLQCD Collaboration~\cite{Kaneko:2021tlw} as well as by future improvements of the precision of the experimental data.

\section{Evaluation of $R(D^*)$ and polarization observables}
\label{sec:RDstar}

By using the unitarity bands of the FFs we can compute the pure theoretical expectation values of the ratio $R(D^*)$, the $\tau$-polarization $P_{\tau}(D^*)$ and the longitudinal $D^*$-polarization $F_L(D^*)$, obtaining
\bea
    \label{eq:Rds}
    R(D^*) & = & 0.275 \pm 0.008 ~ , ~ \nonumber \\[2mm]
    \label{eq:Ptau}
    P_{\tau}(D^*) & = & -0.529 \pm 0.007 ~ , ~ \\[2mm]
    \label{eq:FL}
    F_L(D^*) & = & 0.414 \pm 0.012 ~ \nonumber
\eea
to be compared with the experimental values
\bea
    \label{eq:Rdsexp}
    R(D^*)\vert_{\rm{exp}} & = & 0.295 \pm 0.011 \pm 0.008 ~ \qquad ~ \mbox{Ref.~\cite{HFLAV:2019otj}} ~ , ~ \nonumber \\[2mm]
    \label{eq:Ptauexp}
    P_{\tau}(D^*)\vert_{\exp} & = & -0.38 \pm 0.51^{+0.21}_{-0.16} ~ \qquad \qquad \mbox{Ref.~\cite{Belle:2016dyj}} ~ , ~ \\[2mm]
    \label{eq:FLexp}
    F_L(D^*)\vert_{\exp} & = & 0.60 \pm 0.08 \pm 0.04 ~ \qquad \qquad \mbox{Ref.~\cite{Belle:2019ewo}} ~ . ~ \nonumber
\eea
While the theoretical and the experimental values of $P_{\tau}(D^*)$ are in agreement (mainly due to the larger experimental uncertainty), the compatibility for $R(D^*)$ and $F_L(D^*)$ is at the $\sim1.3 \sigma$ and $\sim 2.1 \sigma$ level, respectively. 
Note that the $R(D^*)$ anomaly results to be smaller with respect to the $\sim 2.5 \sigma$ tension stated by HFLAV Collaboration~\cite{HFLAV:2019otj}.

\begin{figure}[htb!]
\begin{center}
\includegraphics[scale=0.625]{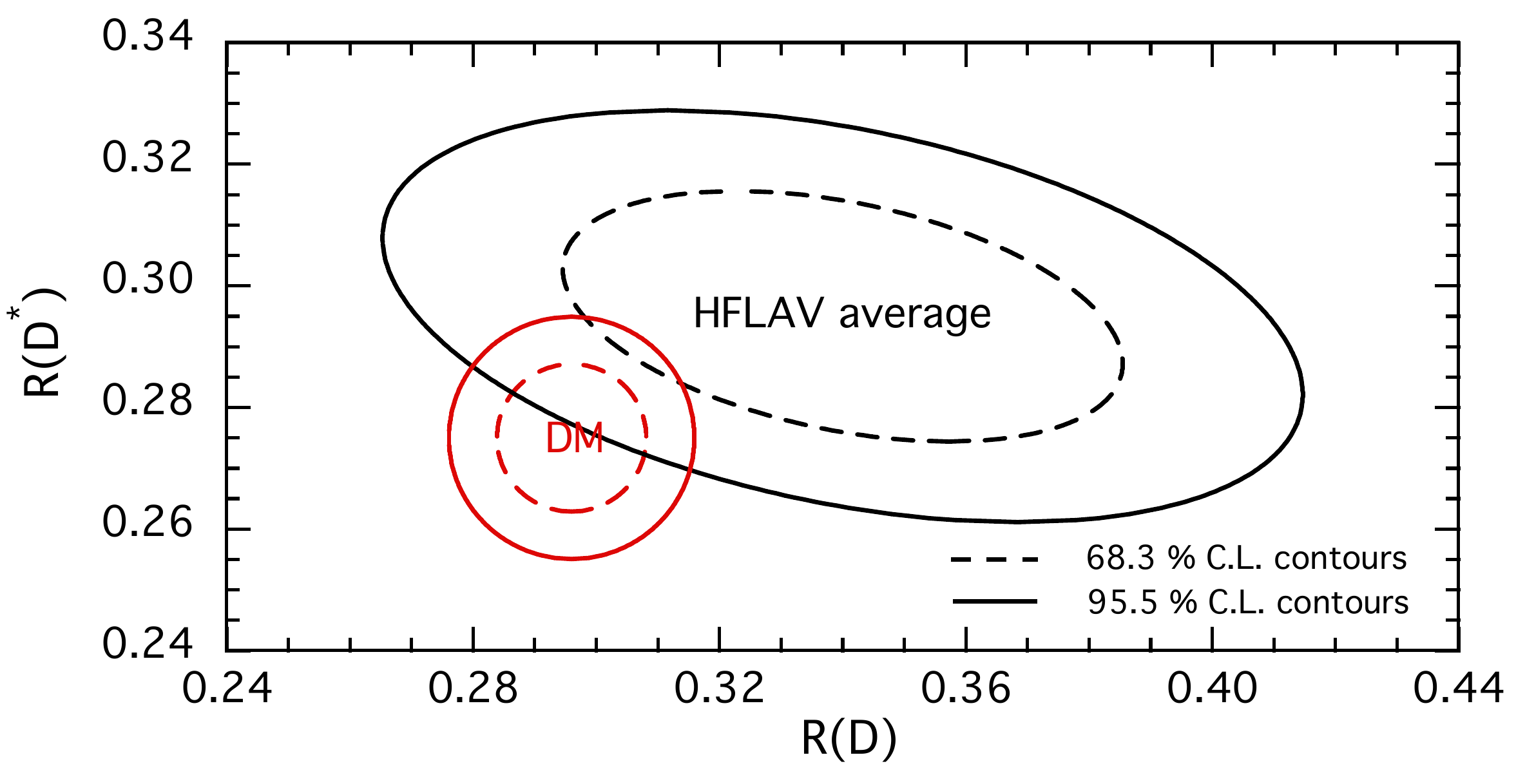}
\caption{\it \small The contour plots of the DM results for the ratios $R(D)$ and $R(D^*)$, obtained respectively in Ref.~\cite{Martinelli:2021onb} and in this work, compared with those corresponding to the experimental world averages from HFLAV~\cite{HFLAV:2019otj}.}
\label{fig:ellipse}
\end{center}
\end{figure}

In Ref.~\cite{Martinelli:2021onb} the DM method was applied to the final lattice data for the $B \to D \ell \nu_\ell$ transition provided by the FNAL/MILC Collaboration~\cite{MILC:2015uhg}. We obtained for the ratio $R(D)$ the pure theoretical estimate $R(D) = 0.296 \pm 0.008$, which is consistent with the experimental world average $R(D)\vert_{\rm{exp}} = 0.340 \pm 0.027 \pm 0.013$~\cite{HFLAV:2019otj} at the $\sim 1.4 \sigma$ level.
In Fig.~\ref{fig:ellipse} we show the comparison of the DM results for the two ratios $R(D)$ and $R(D^*)$ with the corresponding experimental world averages from HFLAV~\cite{HFLAV:2019otj}.
Note that we have considered our values for $R(D)$ and $R(D^*)$ as uncorrelated. This is motivated by the absence of any information about possible correlations among the lattice FFs entering the $B \to D$ and $B \to D^*$ decays and by the fact that the correlation induced by the vector transverse susceptibility $\chi_{1^-}(0)$, which is present in both channels, is very mild, as we have explicitly checked.

\section{Conclusions}
\label{sec:conclusions}

In this work we have applied the DM method~\cite{DiCarlo:2021dzg,Martinelli:2021onb} to the lattice computations of the FFs entering the semileptonic $B \to D^* \ell \nu_\ell$ decays, produced recently by the FNAL/MILC Collaborations~\cite{FermilabLattice:2021cdg} at small, non-zero values of the recoil. 
Thanks to the DM method the FFs have been extrapolated in the whole kinematical range accessible to the semileptonic decays in a completely model-independent and non-perturbative way, implementing exactly both unitarity and kinematical constraints.

Using our theoretical bands of the FFs we have determined $\vert V_{cb} \vert$ from the experimental data and computed $R(D^*)$ from theory.
Our final result for $\vert V_{cb} \vert$ is $\vert V_{cb} \vert = (41.3 \pm 1.7) \cdot 10^{-3}$, which is compatible with the latest inclusive determination $\vert V_{cb} \vert_{\rm{incl}} = (42.16 \pm 0.50) \cdot 10^{-3}$~\cite{Bordone:2021oof} at the $0.5\sigma$ level. 
Moreover, we have obtained the pure theoretical value $R(D^*) = 0.275 \pm 0.008$, which is compatible with the experimental world average at the $\sim 1.3 \sigma$ level.
Together with future improvements of the precision of experimental data, new forthcoming lattice determinations of the FFs at non-zero recoil, expected from the JLQCD Collaboration~\cite{Kaneko:2021tlw}, will be crucial to confirm our present indication of a sizable reduction of the $\vert V_{cb} \vert$ puzzle.

\section*{Acknowledgements}

We acknowledge PRACE for awarding us access to Marconi at CINECA (Italy) under the grant PRA067. We also acknowledge use of CPU time provided by CINECA under the specific initiative INFN-LQCD123. S.S.~is supported by the Italian Ministry of Research (MIUR) under grant PRIN 20172LNEEZ.

\appendix

\section{The DM method}
\label{sec:appA}

In this Appendix we briefly recall the main features of the DM method applied to the description of a generic FF $f(q^2)$ with definite spin-parity.

Let us consider a set of $N$ values of the FF, $\{ f \} = \{ f(z_j) \}$ with $j = 1, 2, ..., N$, where $z$ is the conformal variable
\be
    \label{eq:z}
    z(q^2) \equiv \frac{\sqrt{t_+ - q^2} - \sqrt{t_+ - t_-}}{\sqrt{t_+ - q^2} + \sqrt{t_+ - t_-}} ~ 
\ee
with $t_\pm \equiv (m_{B} \pm m_{D^*})^2$ in the case of our interest and $z_j \equiv z(q_j^2)$.
Then, the FF at a generic value of $z = z(q^2)$ is bounded by unitarity, analyticity and crossing symmetry to be in the range\,\cite{DiCarlo:2021dzg}
\be
  \beta(z) - \sqrt{\gamma(z)} \leq f(z) \leq \beta(z) + \sqrt{\gamma(z)} ~ , ~
    \label{eq:bounds}
\ee 
where 
\bea
      \label{eq:beta_final}
      \beta(z) & \equiv & \frac{1}{\phi(z, q_0^2) d(z)} \sum_{j = 1}^N f(z_j) \phi(z_j, q_0^2) d_j \frac{1 - z_j^2}{z - z_j} ~ , ~ \\
      \label{eq:gamma_final}
      \gamma(z) & \equiv &  \frac{1}{1 - z^2} \frac{1}{\phi^2(z, q_0^2) d^2(z)} \left[ \chi(q_0^2) - \chi_{\{f\}}^{DM}(q_0^2) \right] ~ , ~ \\
      \label{eq:chiDM}
      \chi_{\{f\}}^{DM}(q_0^2) & \equiv & \sum_{i, j = 1}^N f(z_i)f(z_j) \phi(z_i, q_0^2)  \phi(z_j, q_0^2) d_i d_j \frac{(1 - z_i^2) (1 - z_j^2)}{1 - z_i z_j} ~ 
\eea
with
\be
    \label{eq:dcoef}
    d(z) \equiv \prod_{m = 1}^N \frac{1 - z z_m}{z - z_m} ~ , ~ \qquad d_j  \equiv \prod_{m \neq j = 1}^N \frac{1 - z_j z_m}{z_j - z_m} ~ . ~
\ee
In the above Equations $\chi(q_0^2)$ is the dispersive bound, evaluated at an auxiliary value $q_0^2$ of the squared 4-momentum transfer using suitable two-point correlators, and $\phi(z, q_0^2)$ is a kinematical function appropriate for the given form factor\,\cite{Boyd:1997kz}. The kinematical function $\phi$ may contain the contribution of the resonances below the pair production threshold $t_+$.

Unitarity is satisfied only when $\gamma(z) \geq 0$, which implies
\be
    \label{eq:UTfilter}
    \chi(q_0^2) \geq \chi_{\{f\}}^{DM}(q_0^2) ~ . ~
\ee
Since $\chi_{\{f\}}^{DM}(q_0^2)$ does not depend on $z$, Eq.\,(\ref{eq:UTfilter}) is either never verified or always verified for any value of $z$.
This leads to the first important feature of our method: the DM {\it unitarity filter}\,(\ref{eq:UTfilter}) represents a parameterization-independent implementation of unitarity for the given set of input values $\{ f \}$ of the FF.

We point out another important feature of the DM approach.
When $z$ coincides with one of the data points, i.e.~$z \to z_j$, one has $\beta(z) \to f(z_j)$ and $\gamma(z) \to 0$.
In other words the DM method reproduces exactly the given set of data points.
This leads to the second important feature of our method: the DM band given in Eq.~(\ref{eq:bounds}) is equivalent to the results of all possible fits that satisfy unitarity and at the same time reproduce exactly the input data.

The above features may not be shared by truncated parameterisations based on the $z$-expansion, like the Boyd-Grinstein-Lebed (BGL)\,\cite{Boyd:1997kz} fits. Indeed, there is no guarantee that truncated BGL parameterizations reproduce exactly the set of input data and, consequently, the fulfillment of the unitarity constraint may fictitiously depend upon the order of the truncation.

\section{The unitarity filters}
\label{sec:appB}

The non-perturbative values of the dispersive bounds corresponding to the $b \to c$ transition for channels with definite spin-parity have been computed on the lattice at $q_0^2 = 0$ in Ref.\,\cite{Martinelli:2021frl}. 
After subtraction of the contribution of bound states the susceptibilities relevant for the $B \to D^* \ell \nu_\ell$ decays are given by
\bea
      \label{eq:bound1-}
     \chi_{1^-}(0) & = & (5.84 \pm 0.44) \cdot 10^{-4} ~ \mbox{GeV}^{-2} ~ , ~ \\
     \label{eq:bound0-}
     \chi_{0^-}(0) & = & (21.9 \pm 1.9) \cdot 10^{-3} ~ , ~\nonumber  \\     
     \label{eq:bound1+}
     \chi_{1^+}(0) & = & (4.69 \pm 0.30) \cdot 10^{-4} ~ \mbox{GeV}^{-2} ~ ,~ \nonumber
\eea
The kinematical functions associated to the semileptonic FFs reads\,\cite{Boyd:1997kz}
\bea
    \label{eq:phig}
    \phi_{g}(z) & = & 16 r^2 \sqrt{\frac{2}{3\pi}} \frac{(1 + z)^{2}}{\sqrt{1 - z}\left[ (1 + r)(1 - z) + 2 \sqrt{r}(1 + z) \right]^4} ~ , ~ \nonumber \\
    \label{eq:phif}
    \phi_{f}(z, 0) & = & 4 \frac{r}{m_{B}^2} \sqrt{\frac{2}{3\pi}} \, \frac{(1 + z)(1 - z)^{3/2}}{\left[ (1 + r)(1 - z) + 2 \sqrt{r}(1 + z) \right]^4} ~ , ~ \nonumber \\ 
    \label{eq:phiF1}
    \phi_{\mathcal{F}_1}(z, 0) & = & 4 \frac{r}{m_{B}^3} \sqrt{\frac{1}{3\pi}} \frac{(1 + z)(1 - z)^{5/2}}{\left[ (1 + r)(1 - z) + 2 \sqrt{r}(1 + z) \right]^5} ~ , ~ \\
    \label{eq:phiP1}
    \phi_{P_1}(z, 0) & = &  16 \, (1 + r) r^{3/2} \sqrt{\frac{1}{\pi}} \frac{(1 + z)^{2}}{\sqrt{1 - z}\left[ (1 + r)(1 - z) + 2 \sqrt{r}(1 + z) \right]^4} ~ \nonumber 
\eea
with $r \equiv m_{D^*} / m_{B}$.

The presence of resonances below the pair production threshold lead to the following modification of the kinematical function $\phi(z, 0)$\,\cite{Lellouch:1995yv}
\be
    \label{eq:poles}
    \phi(z, 0) \to \phi(z, 0) \cdot \prod_R \frac{z - z(m_R^2)}{1 - z \, z(m_R^2)} ~ , ~
\ee
where $m_R$ is the mass of the resonance $R$.
For the masses of the poles corresponding to $B_c^{(*)}$ mesons with different quantum numbers entering the various FFs we refer to Table III of Ref.~\cite{Bigi:2017jbd}.
We note that in the case of the $B \to D^*$ decay the conformal variable $z$ (see Eq. (11)) ranges from $z(q^2 = t_-) = 0$ to $z(q^2 = 0) = 0.056$ and $z(m_R^2) < 0$. Thus,  the accurate location of the poles in Eq.\,(\ref{eq:poles}) is not crucial.

According to Ref. \,\cite{Boyd:1997kz} there are three unitarity constraints on the FFs of the $B \to D^* \ell \nu_\ell$ decays, namely
\bea
   \label{eq:UTfilter1}
    \chi_{1^-}(0) & \geq & \chi_{\{ g \}}^{DM}(0) ~ , ~ \\
   \label{eq:UTfilter2}
    \chi_{1^+}(0) & \geq & \chi_{\{ f \}}^{DM}(0) + \chi_{\{ \mathcal{F}_1 \}}^{DM}(0) ~ , ~ \\
   \label{eq:UTfilter3}
    \chi_{0^-}(0) & \geq & \chi_{\{ P_1 \}}^{DM}(0) ~ , ~
\eea
where $\chi_{\{ g, f,  \mathcal{F}_1, P_1 \}}^{DM}(0)$ are given by Eq.\,(\ref{eq:UTfilter}) in terms of the known values of the corresponding FF.

\section{Implementation of the kinematical constraints}
\label{sec:appC}

In this Appendix we summarize the basic steps of the procedure adopted to fulfill the KCs\,(\ref{eq:KC1})-(\ref{eq:KC2}) in our sample of events for the FFs.
For ease of presentation we limit ourselves to the KC\,(\ref{eq:KC1}) at minimum recoil $w = 1$, which can be rewritten as
\be
    \label{eq:KC1_app}
    \mathcal{F}_1(1) = \widetilde{f}(1) ~ , ~
\ee
where the FF $\widetilde{f}(w)$ is defined simply as $\widetilde{f}(w) = m_B (1 - r) f(w)$.

We start from a sample of events, each of which is composed by 15 numbers: 12 values of the synthetic lattice points of the FFs (four FFs at three values of the recoil $w$) and the 3 values of the susceptibilities $\chi_{1^-}(0)$, $\chi_{0^-}(0)$ and $\chi_{1^+}(0)$. The sample is generated using a covariance matrix of input data with dimension $15 \times 15$. We consider that the unitarity filters\,(\ref{eq:UTfilter1})-(\ref{eq:UTfilter3}) have been already applied, as described in Section\,\ref{sec:FFs}, so that the sample is composed by events all passing the unitarity filters.

For each event we evaluate Eq.\,(\ref{eq:bounds}) for the two FFs $\mathcal{F}_1$ and $f$ at $w = 1$ (corresponding to $z = 0$), obtaining in this way two lower and two upper bounds, namely $\mathcal{F}_1^{lo(up)}(1)$ and $\widetilde{f}^{lo(up)}(1)$.
Then, we consider only the events for which the dispersive bands for the two FFs overlap each other and for these events we evaluate the intersection between the above bounds to obtain the lower and upper bounds for the common value $\overline{f}(1)$ of the KC\,(\ref{eq:KC1_app}) ,
viz.
\bea
    \overline{f}^{lo}(1) & = & \mbox{max}\left[ \mathcal{F}_1^{lo}(1), ~ \widetilde{f}^{lo}(1) \right] ~ , ~ \nonumber \\[2mm]
    \overline{f}^{up}(1) & =& \mbox{min}\left[ \mathcal{F}_1^{up}(1), ~ \widetilde{f}^{up}(1) \right] ~ , ~ \nonumber
\eea
so that
\be
    \label{eq:range_KC}
    \overline{f}^{lo}(1) \leq \overline{f}(1) \leq \overline{f}^{up}(1) ~ , ~ 
\ee
where $\overline{f}(1) \equiv \mathcal{F}_1(1) = \widetilde{f}(1)$.
As discussed in Ref.\,\cite{DiCarlo:2021dzg}, we consider the value $\overline{f}(1)$ to be uniformly distributed in the range given by Eq.\,(\ref{eq:range_KC}).

We repeat the above steps for all the events of the FFs and, using the subset of events for which the dispersive bands for the two FFs overlap each other, we evaluate the mean values $<\overline{f}^{lo(up)}(1)>$, the standard deviations $\sigma^{lo(up)}$ and the correlations with all the other input data and the one among the bounds, i.e.~$\rho^{lo, up}$.
It follows (see Section V-C of Ref.\,\cite{DiCarlo:2021dzg}) that the mean value and the variance for the FF $\overline{f}(1)$ are given by
\bea
    \label{eq:mean}
    < \overline{f}(1) > & = & \frac{1}{2} \left[ <\overline{f}^{lo}(1)> + <\overline{f}^{up}(1)> \right] ~ , ~ \\[2mm]
    \label{eq:variance}
    \sigma_{\overline{f}(1)}^2 & = & \frac{1}{12} \left[ <\overline{f}^{up}(1)> - <\overline{f}^{lo}(1)> \right]^2 + \frac{1}{3} \left[ (\sigma^{lo})^2 + (\sigma^{up})^2 + 
                                                       \rho^{lo,up} \, \sigma^{lo} \, \sigma^{up} \right] ~ . ~
\eea

Therefore, we increase by one the dimension of the events of our sample by adding the FF $\overline{f}(1)$ at $w = 1$ normally distributed with mean value\,(\ref{eq:mean}) and variance\,(\ref{eq:variance}). Analogously, the dimension of the covariance matrix of the input data becomes $16 \times 16$. The added point is used for evaluating both $\mathcal{F}_1(w)$ and $f(w)$ at a generic value of the recoil $w$.

We repeat the above procedure also for the KC\,(\ref{eq:KC2}) at maximum recoil.

\bibliography{biblio}
\bibliographystyle{biblio}

\end{document}